\documentclass{jpsj-suppl}
\usepackage{txfonts} %Please comment out this line unless the txfonts package is availabe in your LaTeX system.
\usepackage{bm}% bold math
\newcommand{\veps}{\varepsilon}

\title{Novel Linear Algebraic Theory and One-hundred-million-atom Electronic Structure Calculation on The K Computer}

\author{Takeo \textsc{Hoshi}$^{1,2}$, Keita \textsc{Yamazaki}$^{1}$, Yohei \textsc{Akiyama}$^{1}$
}

\inst{$^{1}$Department of Applied Mathematics and Physics, Tottori University, Koyama Minami, Tottori 680-8550, Japan \\
$^{2}$Japan Science and Technology Agency, Core Research for Evolutional Science and Technology (JST-CREST), 5, Sanbancho, Chiyoda-ku, Tokyo 102-0075, Japan
}

\email{hoshi@damp.tottori-u.ac.jp}

\abst{
A novel linear-algebraic algorithm, multiple Arnoldi method, was developed
in an interdisciplinary study between physics and applied mathematics 
and realized one-hundred-million-atom (100-nm-scale) electronic state calculations on the K computer. 
The algorithms are Krylov-subspace solvers for generalized shifted linear equations 
and were implemented in our order-$N$ calculation code ELSES (http://www.elses.jp/).  
Moreover, a method for calculating eigen states is presented as a theoretical extension. 
}

\kword{order-$N$ electronic structure theory,  Krylov subspace, 100-nm-scale materials}

\begin{document}
\maketitle

\section{Introduction}

Large scale electronic structure calculation, with thousand atoms or more, 
is of great importance both in science and technology.
Recently, 
calculations with ten million atoms 
were realized on the K computer 
\cite{HOSHI-mArnoldi, HOSHI-3013-KEI-BENCH}
by our code ELSES (http://www.elses.jp/).
The computational cost is order-$N$ or is proportional to the number of atoms $N$, 
as shown in Fig. 3(a) of Ref.~\cite{HOSHI-mArnoldi}.
The present paper presents more recent calculations 
for larger systems with one hundred million atoms
and the calculations are called  
\lq 100-nm-scale calculation',
because one hundred million atoms are 
those in silicon single crystal with the volume of $V=$(126nm)$^3$. 
The present paper also presents 
a calculation method of specific eigen states
with a polymer example.
The calculation was carried out with modeled 
(tight-binding) electronic structure theory based on {\it ab initio} calculations. 
The detailed theory is described in Ref.~\cite{HOSHI-mArnoldi}

\section{Method and parallel efficiency}

The used linear algebraic method is
called multiple Arnoldi (mArnoldi) method.   \cite{HOSHI-mArnoldi}
The mathematical foundation is 
the \lq generalized shifted linear equation', 
or the set of  linear equations 
\begin{eqnarray}
 ( z S -H ) \bm{x} = \bm{b},
 \label{EQ-SHIFT-EQ}
\end{eqnarray}
where $z$ is a (complex) energy value and 
$H$ and $S$ denote
the Hamiltonian and overlap matrices 
in the linear-combination-of-atomic-orbital (LCAO) representation, respectively.
The matrices are sparse real-symmetric $M \times M$ matrices, 
and $S$ is positive definite.  
The vector $\bm{b}$ is the input and the vector $\bm{x}$ is the solution vector. 
Equation (\ref{EQ-SHIFT-EQ}) is solved, 
instead of the original generalized eigen-value equation 
\begin{eqnarray}
H  \bm{\phi}_k = \varepsilon_k S \bm{\phi}_k.
 \label{EQ-GEV-EQ}
\end{eqnarray}
The method is purely mathematical and 
may be applicable to other problems. 

The mArnoldi method \cite{HOSHI-mArnoldi} 
reduces the problem into a set of many small $(\nu \times \nu)$ standard eigen-value equations 
defined within the Krylov sub(Hilbert)spaces of $\{ {\cal L}_\nu^{(j)} \}$,
where $j$ is the basis index $(j=1,2,3,....M)$
and $\nu$ is the subspace dimension typically $\nu = 30-300$. 
The $j$-th subspace (${\cal L}_\nu^{(j)}$) contains  
the $j$-th unit vector of $\bm{e}_j \equiv (0,0,..1_j, 0,.,0_M)^{\rm t}$
($\bm{e}_j \in {\cal L}_\nu^{(j)}$).
For each subspace eigen-value equation, 
eigen levels $\{ \veps^{(j)}_\alpha \}$ and eigen vectors $\{ \bm{v}^{(j)}_\alpha \}$ 
($\alpha = 1,2,.., \nu$) are obtained and are called 
subspace eigen levels and subspace eigen vectors, respectively. 
The Green's function is determined by the subspace eigen levels and vectors 
and gives the total energy and force. 

\begin{figure}[htbp] 
\begin{center}
  \includegraphics[width=14cm]{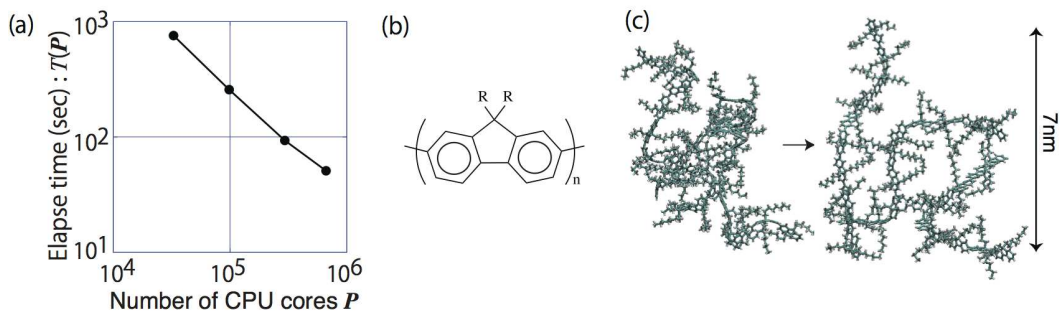}
 %  \vspace{8cm}
\end{center}
%\vspace{-5mm}
\caption{\label{fig-BENCH} 
(a) Parallel efficiency on the K computer with $N=$103,219,200 atoms.
The calculations were carried out with
the total number of processor cores 
$P$ = 32,768, 98,304, 294,912, 663,552 (all cores). 
The calculated material is 
sp$^2$-sp$^3$ nano-composite carbon solid. 
(b) Structure of poly-(9,9 dioctyl-fluorene).  
Here $R \equiv$ C$_8$H$_{17}$. 
(c) A molecular dynamics simulation of a poly-(9,9 dioctyl-fluorene) system.
}
\end{figure}

Figure \ref{fig-BENCH}(a) shows 
the parallel efficiency on the K computer
with one hundred million atoms ($N=$103,219,200).
The elapse time  $T$ is measured as the function of 
the number of used processor cores $P$ $(T=T(P))$,
where 
$P = P_{\rm min} \equiv 32,768$, $P_{a} \equiv$98,304, $P_{b} \equiv$294,912, $P_{\rm all} \equiv 663,552$ (all cores). 
The resultant benchmark is called 
\lq strong scaling' in the high-performance computation society.
The calculated material is 
sp$^2$-sp$^3$ nano-composite carbon solid. \cite{HOSHI-3013-KEI-BENCH}
The time of the total energy calculation was  measured for a given atomic structure.
The measured parallel efficiency, $\alpha(P) \equiv T(P)/T(P_{\rm min})$, was
$\alpha(P_{a})=0.98$, 
$\alpha(P_{b})=0.90$ and 
$\alpha(P_{\rm all})=0.73$.

%%%%%%%%%%%%%%%%%%%%%%%%%%%%%%%%%%%%%%%%%%
%%%%%%%%%%%%%%%%%%%%%%%%%%%%%%%%%%%%%%%%%%

\section{Method for calculating eigen states}

\begin{figure*}[htbp] 
\begin{center}
  \includegraphics[width=15cm]{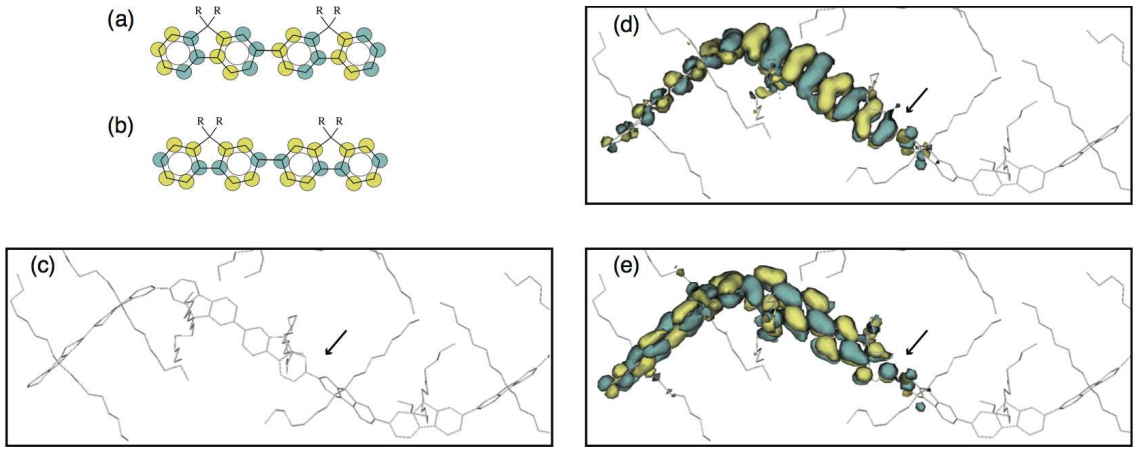}
\end{center}
%\vspace{5mm}
\caption{\label{fig-PF2076atoms-HOMO} 
(a)(b)  Schematic figures of
the HO or LO ($\pi$) state  of fluorene dimer, respectively.  
(c)-(e) A part of the poly-fluorene system is shown in (c) 
and the calculated  HO and LU wavefunctions are shown in (d) and (e), respectively.
Only carbon atoms and carbon-carbon bonds are drawn.
}
\end{figure*}

This section is devoted to a calculation method for individual eigen states in the mArnoldi method, 
since it is missing in our previous papers. \cite{HOSHI-mArnoldi,HOSHI-3013-KEI-BENCH}
A conjugated polymer system depicted in Figs.~\ref{fig-BENCH}(b)(c),
poly-(9,9 dioctyl-fluorene), was chosen as a test system.
The poly-fluorene and its family are famous 
as a hopeful candidate for industrial lighting applications. 
A previous theoretical paper \cite{ZEMPO-JPCM-2008} investigates 
related small molecules 
and drove us to calculations of polymer materials with non-ideal (amorphous-like) structures.
The present calculations are intended for a methodological test, in particular, on $\pi$ states.
As a reference data, 
the monomer and dimer were calculated 
by the present method and 
the {\it ab initio} calculation of Gaussian$^{\rm (TM)}$
and the results agree reasonably among the two methods. \cite{HOSHI-mArnoldi}  
The calculated highest-occupied (HO) and lowest-unoccupied (LU) states  of dimer
are contributed only by  the $\pi$ states of benzene rings, as in benzene, 
and are illustrated in Fig.\ref{fig-PF2076atoms-HOMO}(a) and (b), respectively. 
A molecular dynamics simulation with 2076 atoms was carried out 
as shown in Fig.~\ref{fig-BENCH}(c).
An artificially packed structure was thermally relaxed 
within the period of $T$=80 ps, 
so as to obtain non-ideal polymer structures.
~\cite{HOSHI-mArnoldi}
The calculated DOS is shown in
Fig.2 of Ref.~\cite{HOSHI-mArnoldi}. 
Hereafter the same poly-fluorene system will be discussed
in  the same calculation conditions as  in the DOS. 

The method for calculating the $k$-th eigen level ($\varepsilon_k$) and vector ($\bm{\phi}_k$) are explained; 
The Green's function gives 
the integrated density of states $n=n(\eta)$ and its inverse function $\eta = \eta(n)$.
The $k$-th eigen level is determined in the inner product form of
$\varepsilon_k : = \eta(k-1/2)$
(See Footnote 9 of Ref.~\cite{HOSHI-mArnoldi}).
Then, the $k$-th eigen vector ($\bm{\phi}_k$) is calculated from
the subspace eigen vectors of which energy levels lie 
in the range of $\eta(k-1) < \varepsilon < \eta(k)$ so
that the $j$-th component $(\bm{e}_j^{\rm t} \bm{\phi}_k)$ is determined  as
\begin{eqnarray}
\bm{e}_j^{\rm t} \bm{\phi}_k : = 
\sum_{\alpha} \bm{e}_j^{\rm t} \bm{v}_{\alpha}^{(j)} \int_{\eta(k-1)}^{\eta(k)}  
\delta (\varepsilon - \varepsilon_{\alpha}^{(j)}) d \varepsilon.
\label{EQ-EIGEN-KRY}
\end{eqnarray}
Here the delta function in the above equation is a \lq smoothed' one. 
Additional techniques for calculating eigen vectors are given in
in Appendix.

Figure \ref{fig-PF2076atoms-HOMO}(c) is a part of the polymer system
at which  the calculated HO and LU wavefunctions are localized, 
as in Figs \ref{fig-PF2076atoms-HOMO}(d) and (e), respectively. 
They are $\pi$ states and two features are found; 
(I) The characteristic node structures are observed and
are the same as in the dimer case (Fig.~\ref{fig-PF2076atoms-HOMO}(a) and (b)). 
In short, 
the HO and LU wavefunctions of the polymer are connected HO and LU wavefunctions of dimer, respectively. 
(II) The wavefunctions are localized on the chain structure among four monomers and terminated 
at one inter-monomer boundary indicated by an arrow.
At that boundary, the structure is largely tilted (or twisted) and the $\pi$ states are disconnected. 
There features are confirmed in the wavefunctions given by the exact diagonalization method.

%%%%%%%%%%%%%%%%%%%%%%%%%%%%%%%%%%%%%%%%%%
%%%%%%%%%%%%%%%%%%%%%%%%%%%%%%%%%%%%%%%%%%

\section{Summary and future outlook}

The present paper shows methods and results of 
one-hundred-atom electronic structure calculations
based on our novel linear algebraic algorithm. 
For future outlook, 
methods for automatic determination of 
model (tight-binding) parameters 
are developing, \cite{NISHINO-2013} so as to enhance studies of various materials.
Matrix data $(H,S)$ for materials 
\cite{ELSES-MATRIX-LIBRARY} and
a small application for matrix solvers \cite{EIGEN-TEST} are 
prepared
for further interdisciplinary study between physics and applied mathematics.

\vspace{3mm}

{\bf Acknowledgments} \hspace{3mm}
This research is partially supported by Grant-in-Aid 
for Scientific Research
(Nos. 23540370 and 25104718)
from the MEXT of Japan. 
The K computer was used 
in the research proposals of hp120170 and hp120280.
Supercomputers were also used 
at the Institute for Solid State Physics, University of Tokyo, 
at the Research Center for 
Computational Science, Okazaki,
and at the Information Technology Center, University of Tokyo.
The authors thank Y. Zempo (Hosei University) and M. Ishida (Sumitomo Chemical Co., Ltd) 
for providing the structure model of poly-fluorene system.
The wavefunctions in Figs.~\ref{fig-PF2076atoms-HOMO}(d)(e) are drawn 
by an original python-based visualization tool VisBAR\_wave\_batch, 
a part of VisBAR package. \cite{HOSHI-3013-KEI-BENCH}

%\end{acknowledgments}

\appendix

\section{Refinement techniques for specific eigen vectors \label{SEC-CALC-EIGEN}}
Refinement techniques for  a specific eigen vector  ($\bm{\phi}$) is explained,
where the level index ($k$) is discarded. 
A subspace eigen vector ($\bm{v}_{\alpha}^{(j)}$) has 
the sign-flipping freedom ($\bm{v}_{\alpha}^{(j)} \Rightarrow - \bm{v}_{\alpha}^{(j)}$)
and  the correct sign should be assigned. 
In the code, 
a \lq reference' subspace eigen vector, denoted as $\bm{v}_{{\rm ref}}$, 
is chosen among the subspace eigen vectors (\{ $\bm{v}_{\alpha}^{(j)}$ \})
so that the \lq reference'  subspace eigen vector has the largest weight contribution to the eigen vector $\bm{\phi}$.
The sign of a subspace eigen vectors 
($\bm{v}_{\alpha}^{(j)}$) in Eq.~(\ref{EQ-EIGEN-KRY}) is determined
so that the inner product between $\bm{v}_{{\rm ref}}$ and $\bm{v}_{\alpha}^{(j)}$ is positive 
($\bm{v}_{{\rm ref}}^{\rm t} \bm{v}_{\alpha}^{(j)}>0$).
If the subspace eigen vectors are exact, 
the above procedure gives the correct sign of wavefuction. 
Since this method, called inner product method, 
is insufficient in several cases, 
a further refinement technique, called local flipping method', is used.
The target eigen level and vector are denoted as
$\varepsilon$, and $\bm{\phi} \equiv (a_1, a_2, ..... a_M)^{\rm t}$.
The method gives
an iterative local sign flipping processes $(a_l \Rightarrow - a_l)$,
so as to reduce the residual norm ${\cal R} \equiv \bm{r}^{\rm t} \bm{r}$
defined by the residual vector of  
$\bm{r} \equiv (H - \varepsilon S) \bm{\phi}$.
The details will appear elsewhere. 
As an additional technique for faster convergence, 
the LU state $\bm{\phi}$
(Fig.~\ref{fig-PF2076atoms-HOMO}(e))
was calculated after the calculation of the HO state $\bm{\phi}_{\rm HO}$
(Fig.~\ref{fig-PF2076atoms-HOMO}(d))
with the redefinition of the residual norm $R$ as
${\cal R} \equiv \bm{r}^{\rm t} \bm{r}
+ c \, | \bm{\phi}_{\rm HO} ^{\rm t} S \bm{\phi} |^2 
 \label{EQ-LFP-DEF-RN2}$
with a positive parameter $c$(=5a.u.).
The minimization of  the additional term is 
an effective orthogonality constraint 
$(| \bm{\phi}_{\rm HO} ^{\rm t} S \bm{\phi} | \rightarrow 0)$.
The wavefunctions in Fig.~\ref{fig-PF2076atoms-HOMO}(d)(e)
are obtained after $10^1$-$10^2$ local sign flipping iterates.
The above sign flipping procedure consumes
only a small computational cost, 
since no matrix-vector multiplication is required. 
A further refinement can be achieved by another method, 
such as Jacobi-Davidson method \cite{TEMPLATE}, 
though it requires matrix-vector multiplications.

\end{document}